\documentclass[iop]{emulateapj}

\usepackage{graphicx}
\usepackage{color}

\newcommand{\msun}{$M_{\sun}$}
\newcommand{\hst}{{\it HST}}
\newcommand{\ha}{H$\alpha$}

\slugcomment{To appear in ApJL}

\shorttitle{Star clusters in the nuclear ring of NGC~2328}
\shortauthors{V\"ais\"anen et al.}

\begin{document}



\title{Star clusters in a nuclear star-forming ring: The disappearing string of pearls}


\author{Petri V\"ais\"anen\altaffilmark{1,2}, Sudhanshu Barway\altaffilmark{1},
Zara Randriamanakoto\altaffilmark{1,3}}
\affil{$^1$South African Astronomical Observatory, P.O. Box 9 Observatory, Cape Town, South Africa}
\affil{$^2$Southern African Large Telescope, P.O. Box 9 Observatory, Cape Town, South Africa}
\affil{$^3$University of Cape Town, Astronomy Department, Private Bag X3, Rondebosch 7701, South Africa}
\email{petri@saao.ac.za}

\begin{abstract}

An analysis of the star cluster population in a low-luminosity early type galaxy, NGC~2328, is presented. 
The clusters are found in a tight star-forming nuclear spiral/ring pattern and we also identify a bar from
structural 2D decomposition.
These massive clusters are forming very efficiently in the circum-nuclear environment, they are young, possibly all less than 30 Myr of age.  The clusters indicate an azimuthal age gradient, consistent with a "pearls-on-a-string" formation scenario suggesting bar driven gas inflow.
The cluster mass function has a robust down-turn at low masses at all age bins.  Assuming clusters are born with a power-law distribution, this indicates extremely rapid disruption at time-scales of just several Myr.  If found to be typical, it means that clusters born in dense circum-nuclear rings do not survive to become old globular clusters in non-interacting systems. \\

\end{abstract}


\keywords{galaxies: evolution --- galaxies: individual(NGC 2328) --- galaxies: star clusters: general --- galaxies: starburst}

\section{Introduction}

Young massive star clusters (YMC), or super star clusters (SSC), have been found to be ubiquitous in interacting gas rich galaxies and dwarf starbursts \citep{por2010}. There are not many studies of YMCs in early type galaxies (ETG), though when they are found they tend to inhabit nuclear star-forming regions \citep[e.g.][]{buta2000}.  YMCs thus appear to form in high-pressure environments - but they may also dissolve quickly there \citep[e.g.][]{kruijssen2014}.  Which clusters survive to perhaps become old globular clusters, and in which conditions?

Here we present a study of the cluster population in NGC\,2328, a little-studied low-luminosity (M$_V ^{tot} = -18.5$ mag) galaxy at (the adopted) 18 Mpc distance with an S0 classification.  In the centre of the largely featureless isolated galaxy, \hst\ images reveal a tight and bright nuclear spiral, or a ring, of $\sim200$ pc radius. The feature resolves itself into numerous "hot spots", point sources which are candidate star clusters.  At this distance 0.1\arcsec, just a little larger than the PSF, corresponds to  
$\sim 9$~pc, which is larger than the expected size of YMCs.   Other studies have found both young and intermediate age clusters in star-forming rings \citep[e.g.][]{buta2000,maoz2001}, 
while recently \citet{degrijs2013} demonstrated very rapid disruption in a starburst ring of a spiral galaxy.  The aim of this study is to define the age and mass distribution of YMCs in NGC\,2328 to explore the effect of the ring-environment on star clusters.

\section{Data}

We made use of archival HST imaging data for NGC\,2328, downloaded from The Hubble Legacy Archive (HLA)\footnote{http://hla.stsci.edu/hlaview.html} at the Space Telescope Science Institute and were processed by the HLA pipeline (see Table~\ref{table1} for details). The $H$ and \ha\ images were aligned to the $V$ and $I$ bands which already were aligned.
The $H$-band image with a native pixel scale of 0.09" was resampled to a common scale of 0.05''/pixel with the optical data, so that images in all bands were aligned to $\sim1/4$ pixel accuracy. The FWHM of the PSFs were $\sim0.08$" and 0.15" in the optical and NIR data, respectively.

\begin{deluxetable}{llrrr}
\tabletypesize{\small}
\tablewidth{0pt}
\tablecaption{\hst\ data for NGC\,2328.}
\tablecolumns{2}
\tablehead{
\colhead{Instrument} &
\colhead{Filter} &
\colhead{Exp.\ time} &
\colhead{Prop.\ ID} &
\colhead{PI} \\
\colhead{} &
\colhead{} &  
\colhead{Sec} &
\colhead{} &
\colhead{} 
}
\startdata
WFPC2/PC & F555W ($V$)  &    160   & 5999 & A. Phillips \\
WFPC2/PC & F814W ($I$)       & 320   & 5999 & A. Phillips \\
WFPC2/PC & F658N (H$_{\alpha}$)  &  4400  & 6785 & M. Malkan \\
WFC3    & F160W ($H$)     & 394 & 11219 &  A. Capetti 
\enddata
\label{table1}
\end{deluxetable}

\begin{figure}
\resizebox{0.66\hsize}{!}{\rotatebox{0}{\includegraphics{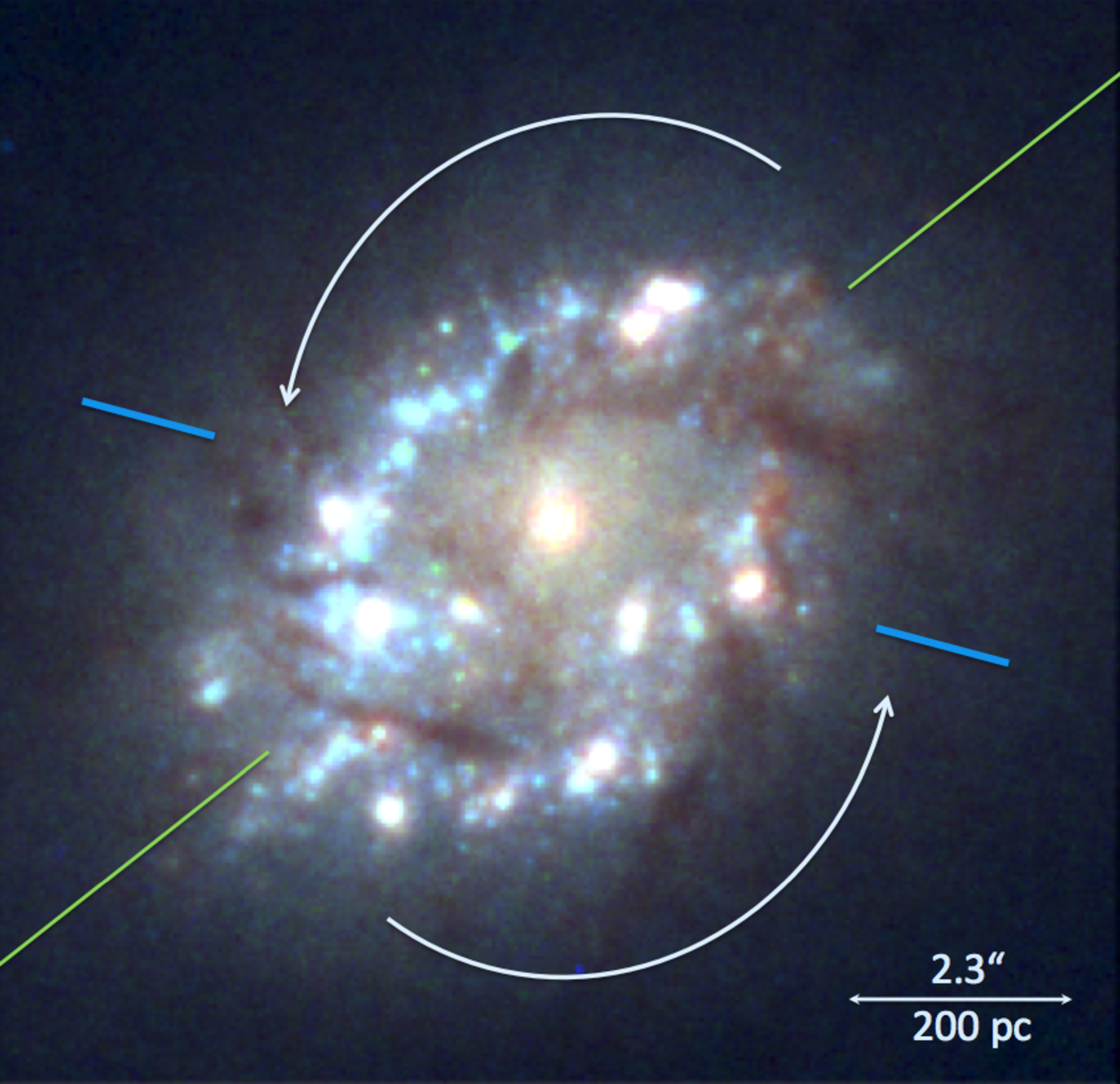}}}
\caption{A 3-color $VIH$ \hst\ view of NGC\,2328, North is up, East is left.  The young star clusters in the tight star-forming ring (or spiral) pattern have an azimuthal age gradient (see Section~\ref{agegrad}). YMCs get younger along the direction of the arrows with the short markers indicating the approximate location of the youngest populations, while the longer markers show the PA of a bar.} 
\label{threecolor}
\end{figure}

We obtained low and medium resolution long-slit spectra of the galaxy in the 3700 to 6900\AA\ wavelength range with the Southern African Large Telescope (SALT, \citet{dod2006}).  
The spectra are seeing limited and do not sample the galaxy at the resolution of individual YMCs, but they provide useful constraints for age modelling in this Letter.  The spectra are described and analysed in detail in a parallel work characterizing the galaxy as a whole (V\"ais\"anen et al, in prep.), here we make use of specific end results regarding metallicity, extinction and rotational velocity, summarized in Table~\ref{table2}.

\subsection{Galaxy structure}

NGC\,2328 is classified as S0 in the RC3 but (RÕ)SBA0 in NED.  
We derived structural parameters from the WFC3 $H$-band image using GALFIT \citep{peng2002}.  A combination of Sersic, Exponential and Gaussian functions gives the best-fit model. The results, given in the Table~\ref{table2}, indicate that NGC\,2328 has no bulge, just a disk component with a bar and an additional point-source in the center.

\begin{deluxetable}{lr}
\tabletypesize{\small}
\tablewidth{0pt}
\tablecaption{Structural and stellar population parameters for NGC\,2328 from $H$-band \hst\ and SALT data, respectively.}
\tablecolumns{2}
\tablehead{}
\startdata
Sersic (represents bar)  &  \\
Total magnitude m$_T$   &  12.04 \\
Effective radius R$_e$  pc &  279 \\
Sersic index $n$        &  0.29 \\
Position angle          & -51.0  \\
Ellipicity $\epsilon$   &  0.30 \\
\hline
Exponential (represents disk)  &  \\
Total magnitude m$_T$   &  10.53 \\
Disk Scale R$_d$ pc       &  545 \\
Position angle          & -66.1 \\
Ellipicity $\epsilon$   &  0.33 \\
\hline
Gaussian (represents point source)  &  \\
Total magnitude m$_T$   &  16.38 \\
FWHM  pc                  &   26 \\
\hline
Metallicity $[Fe/H]$    & 0.008 \\
A$_V$ mag             & 0.8 \\
V$_{rot}$  km s$^{-1}$   & 36 
\enddata
\label{table2}
\end{deluxetable}

\subsection{Photometry of clusters}
\label{photometry}

For detection of point sources in the \hst\ images we used SExtractor \citep{bertin1996} on the $I$-band image which was first un-sharp masked to highlight sources in the complex background of the ring.  Comprehensive tests were performed to balance extraction parameters to detecting the faintest ones without yet introducing clearly spurious sources. We used a similar strategy for SSC detections in \citet{zara2013a}.  Photometry itself was conducted on the original aligned broad-band images using fixed small apertures of 2 pixel (0.1\arcsec) radius, with a sky-annulus from 2.5 to 4 pixel radii.   The diameter of the photometric aperture thus corresponds to about 17 pc physical size.  Aperture corrections were adopted from \citet{dolphin2009} with effects of the re-sampling, and a larger PSF, on the $H$-band image determined using isolated point sources.  All photometry was transformed to the Vega-system \citep{dolphin2009}.  The final photometric catalog at this point consists of 193 cluster candidates. 
Completeness limit Monte Carlo simulations were run and 90\% completeness is still recovered within the ring region at $I\approx21.8, V\approx22.0$ mag or $M_I \approx -9.5, M_V \approx -9.3$.  
We note that individual blue supergiants may still just be seen as individual sources at this limit, but they would not be numerous enough to make any statistical significance \citep[e.g.][]{maoz2001,chandar2010a} in our results which range to $M_V \approx -13.5$ mag.  The presence, or not, of individual red supergiants (RSGs) within a cluster, on the other hand, may have significant stochastic effects on the brightness of an YMC \citep[e.g.][]{degrijs2013,anders2013} especially in the NIR, and we explore this aspect more in our parallel NGC\,2328 work.
  
The \ha\ image was treated separately. It is difficult to uniquely assign \ha\ flux to a given YMC point source because of varying morphologies.  At young ages the gas would be expected to be within the cluster but it will gradually blow out into a growing bubble and finally disperse altogether \citep[see e.g.][]{whitmore2011}.  A further complication arises from any diffuse gas emission in the general region of a given YMC not necessarily related to that one.  We did not use \ha\ fluxes, where the decision on how exactly the sky region is defined affects results. Rather, after experimentation, we noted that the Equivalent Width (EW) of \ha\ {\em within} the 2\arcsec photometric aperture gives more robust results \citep[see also][]{cresci2005}-- essentially, the size of any \ha\ {\em bubble} does not come into play.  EW(\ha) decreases monotonically  with age and is easy to compare to model predictions.  

The EW(\ha) values per YMC candidate were obtained in the following fashion.  The narrow-band  $(NB)$  F658N filter image encompasses \ha\ from the galaxy, and N[II] just at the edge of the  filter transmission.  We used the $I$ broad band $(BB)$ image to obtain a continuum map at the NB wavelength location by finding an optimal scaling for it so that a $NB - f \times BB$ subtraction removes the half dozen stars evident in the NB  frame resulting in a line emission $(LE)$  frame only.  Then, $LE / (f \, BB) \times W(NB) $, where the last term is the width of the $NB$ filter (in \AA) gives an estimate of the EW of \ha\ emission.  See \citet{spector2012} for a detailed discussion of this method, and of inherent assumptions.  We do not correct for N[II] contribution to the NB filter which we estimate to be at most 10\% based on the SALT spectra and the filter transmission curve.
The approach is satisfactory since we do not use the EW values alone on age estimates, but as one out of four independent values in the cluster age and mass modelling. The average YMC EW(\ha) 
is consistent with SALT spectra values.

Finally, we made an empirical nebular emission correction to the $V$ and $H$-band magnitudes using the $LE$ image derived above. Optical line ratios were first measured from the SALT spectra.  Knowing the H$\alpha$+[NII] flux in the NB filter, using the filter transmission curves and line ratios we can directly calculate also the emission line contribution to the $V$ filter. We then assume the \ha\ image describes the H$\beta$+[OIII] emission morphology as well and scale the $LE$ image appropriately before subtracting from $V$-band.
Based on results of \citet{anders2003} we estimate the contribution of emission line flux in $H$-band to be roughly half of the emission line contribution to $V$-band, and made this subtraction as well.  The corrections can {\em brighten} the YMC when there is more emission in the sky annulus than in the target aperture.  Typical differences between the corrected and uncorrected magnitudes are $\sim0.1$ mag in $V$, while $\Delta {\rm mag}$ values in the 0.2 to 0.6 range are found for some 20 sources in the strongest H$\alpha$ emission areas.   

We selected sources with photometric errors $<0.5$ mag in $V,I$ bands from the catalog of 193 sources produced above resulting in 159 cluster candidates.  We checked their FWHMs, the great majority are clustered around the expected PSF size of $\sim1.8$ pixels, the median being 1.9.  The $\sim$10\% of the candidates with more than double this value were checked individually - some of them may be blends or cluster complexes though most would have wider PSFs merely due to the difficulty of accurate FWHM determinations in the complex background.  Since a robust discriminator between the two cases is difficult to define, we did not exclude any of them from the catalog.  However, we checked that ignoring either all, or none, of the wider FWHM cases will not change any results.  We did not set a formal limit in $H$ mag error since we wanted an optically selected candidate list.  Marginal NIR detections are thus included, but the $H$-band role of these cases is diminished due to inverse weighting by error. Results were checked not to change using the 129 sources resulting from setting $H$-error to $<0.5$ mag, or by selecting stricter error cut-offs in all bands for that matter. 

\section{Ages and masses of the YMCs}

It is well known that using only 3 broad band filters for mass/age modelling fails to break degeneracies between age, metallicity and extinction \citep[e.g.][]{anders2004}.  Especially ages between $\sim10$ and $\sim100$ Myr would be impossible to differentiate only with three band data.
Constraining both metallicity and extinction with the SALT spectra (values in Table~\ref{table2}) at the ring locations helps, and
most importantly we made use of \ha\ data.

$V$, $I$ and $H$ band Starburst99 model (\citet{sb99}; Padova tracks, Kroupa IMF, using Z=0.008) magnitudes, which do {\em not} include nebular emission (though nebular continuum is included), were used as well as EW(\ha) values.  The $\chi^{2}$ minimization fitting to these four points used the formal uncertainties from the photometry. $A_V$ was left as a free parameter, but restricted to range $\pm0.5$ mag around 0.8 mag suggested by the spectra. The average of the output was $A_V\approx0.9$.  The masses were calculated based on the average of $V$ and $I$ luminosities after best-fit ages and extinctions were determined. The $H$-band luminosity was not used for mass determination since it is more susceptible to stochastic effects due to red supergiants \citep{gazak2013,degrijs2013}.

\begin{figure*}
\resizebox{0.95\hsize}{!}{\rotatebox{0}{\includegraphics{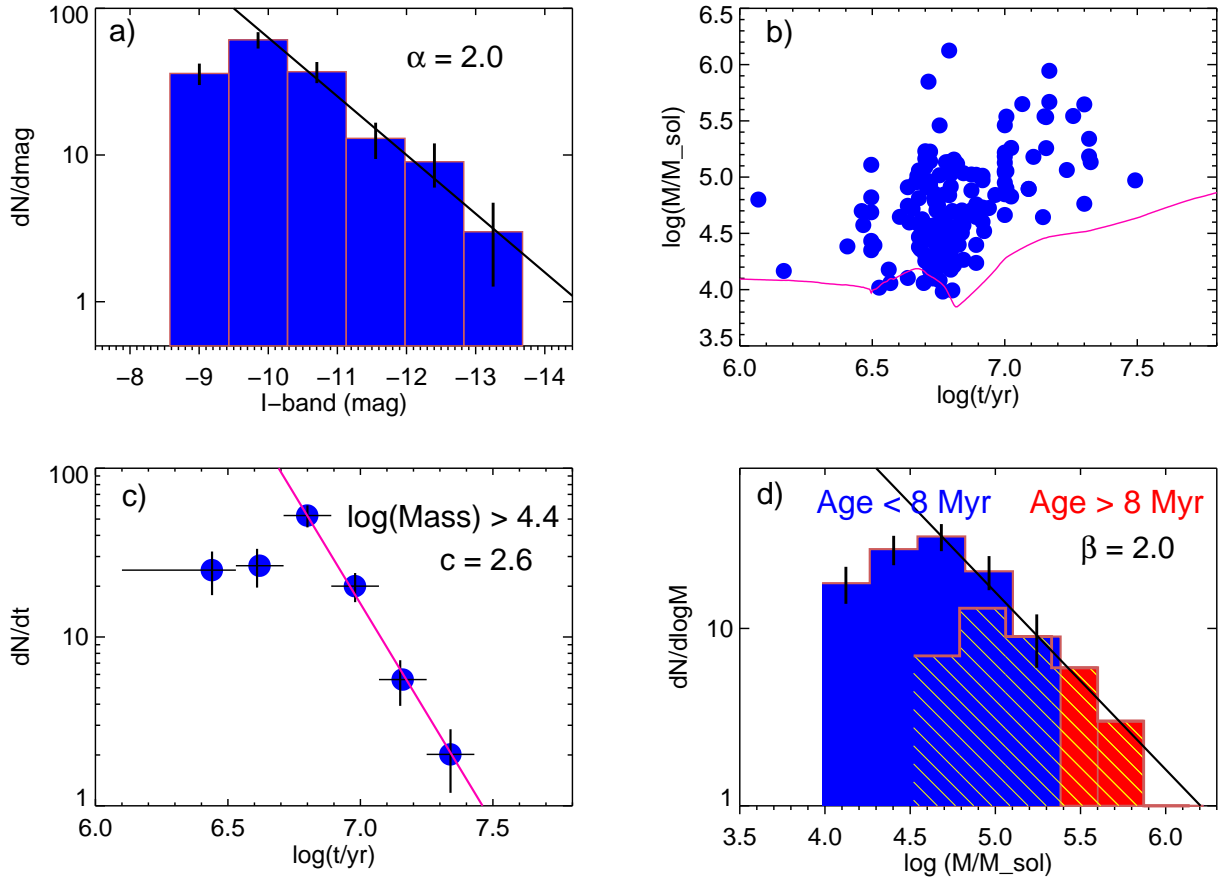}}}
\caption{{\it Top left:}  The $I$-band LF of the YMCs with a canonical  $\alpha=2$ power-law slope.  Photometric incompleteness is expected to set in at $M_I\sim -9.5$ mag, corresponding to the turnover.  {\it Top right:}  The mass vs.\ age  of all the YMCs.  The expected mass of a cluster at the 90\% completeness limit, assuming $A_V=0.8$ mag, is shown as the magenta curve.  {\it Bottom left:}  The age distribution of the clusters above the indicated mass cutoff, with a power-law distribution of the oldest bins also shown.   {\it Bottom right:}  The MF of YMCs below and above the age of 8 Myr are shown as the two histograms. A curve corresponding to a power-law slope of $\beta = 2$ is overplotted.  There is a clear turn-over of the MF in both age bins.}
\label{distributions}
\end{figure*}

The clusters in NGC\,2328 are young, ranging from a few Myr up to 30 Myr.  Figure~\ref{distributions} summarises the luminosity function (LF), the mass function (MF), and also the age distribution of the detected sources.  There is no evidence of any older population. Two key results are the apparent turnover of the MF (Fig.~\ref{distributions}d), and an azimuthal age gradient along the ring (Fig.~\ref{azdist}), as discussed in Section~\ref{discussion}.  We checked that the general characteristics of the mass and age distributions remain the same under a wide variety of modelling choices, including the input models themselves, making the main results robust.  While the \ha\ data drive the age determination they are much less critical for the YMC masses which come out similar with or without \ha, or with different codes (we experimented also with GALEV \citep{kotulla2009} and Yggdrasil \citep{zack2011}).

We note that some fraction of the cluster candidates might be cluster complexes or blends of several SSCs.  The surface density of YMCs in the ring area is about 600 kpc$^{-2}$, approaching confusion limits.  However, with the small aperture sizes used, $r\approx9$ pc, crowding effects are not expected to significantly bias the statistical results  \citep{zara2013a}.  The reasonably small apertures also help in mitigating filter-dependent resolution effects \citep{bastian2014}.

\section{Discussion}
\label{discussion}

\subsection{The turnover of the mass function}

Figure~\ref{distributions}a shows the luminosity function of all our YMCs fitted with $\Phi(L) dL \propto L^{-\alpha} dL$. The slope has a very robust $\alpha \sim 2$ until the 90\% completeness limit (Section~\ref{photometry}) in agreement with "canonical" slopes for nearby star forming galaxies.  The masses vs.\ age of the clusters is then shown in panel b, with the mass limit vs.\ age corresponding to the photometric  completeness indicated.
The age distribution of the YMCs (panel c) at a complete mass range log(\msun)$>4.4$, when fit with $dN / d log(t) \sim t^{-c}$, gives $c \sim -2.6$ for $>5$ Myr. This distribution is a complex function of cluster fading, dissolution, initial formation power-laws, and detection limits \citep[e.g.][]{boutloukos2003,gieles2007,chandar2010b}, and it is not obvious whether an analysis in the narrow age range we find is appropriate. Nevertheless, we note that the very steep slope is indicative of significant dissolution of clusters (dropping \ha\ and deriving older ages still results in $c\sim1.5$).

The mass function is the most interesting distribution. The bright side of the MF in Fig.~\ref{distributions}d  shows typical  $\beta \sim 2$ in a $dN / dM \propto M^{-\beta}$ fit.  Though $\beta$ varies by at least $\pm0.5$ depending on various binning and modelling choices,
the MF {\em turnover} is very robust, whether all or sub-populations of ages are plotted.  The blue histogram shows the youngest $<8$~Myr population, for example.  Given the 90\% completeness limit corresponds to $\sim10^4$\msun\ at the relevant ages the turnover cannot be due to photometric incompleteness only.  A natural explanation of this behaviour is a rapid destruction of clusters at low masses -- {\em assuming} that clusters are born in a power-law distribution, as widely accepted from observations of GMCs and open clusters in our own Milky Way, and more massive young clusters in the Local Group and beyond. 

The result from this analysis thus is that while the more massive clusters appear to have typical YMC characteristics, the less massive cluster population in NGC\,2328 has had to disappear in a very short timescale of 5 to 30 Myr given the photometric depth of our images.  A similar results was found by \citet{degrijs2012}. This would also indicate that such clusters do not typically survive to become old globular clusters -- YMCs thrown out of the densest birth regions by e.g.\ interactions might have a better chance of survival  \citep{kruijssen2014}.
 
\subsection{Age gradient}
\label{agegrad}

\begin{figure}
\resizebox{1.05\hsize}{!}{\rotatebox{0}{\includegraphics{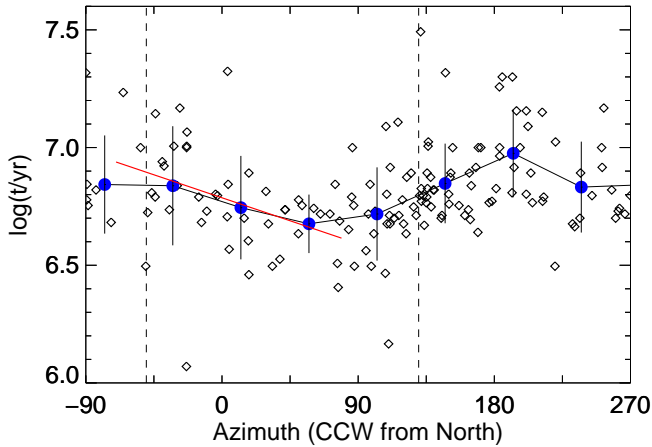}}}
\caption{The ages of YMCs vs. AZ, counter-clockwise along the rotational direction.  The blue points depict the averages in 45 deg bins with the standard deviation of ages in each bin plotted as the uncertainty. The red curve shows a linear fit to the individual points in the $AZ=-70$ to 80 deg range. The dashed vertical lines indicate the bar positions.  See Fig.~\ref{threecolor}} 
\label{azdist}
\end{figure}

Figure~\ref{azdist} plots the Azimuth (compare to Fig.~\ref{threecolor}) of each YMC against their age revealing a possible non-random distribution.  It appears there is a minimum at $AZ\sim60$ to 90, and another one somewhere in-between AZ=230 and 270. While systematic trends over the whole system are somewhat speculative given the uncertainties, for example the youngest and oldest 45-degree bins are just over $1\sigma$ away from the mean, a gradient can be defined better in a section of the ring:  The red curve in Fig.~\ref{azdist} shows a fit to the ages of {\em individual cluster points} over most of the North-East arm and the resulting slope has an uncertainty $3\sigma$ away from a flat line.  The PA of the identified bar is marked and it appears that the youngest SSCs are located some $\sim50$ deg upstream from the points where the bar crosses the rotating ring, while the oldest areas correspond approximately to the ends of the two nuclear spiral arms.
  \citet{mazzuca2008} find, from a sample of several nuclear rings, that the youngest HII regions tend to be correlated with {\em contact points} of the ring and bar induced flows they define to be perpendicular to the bar PA. Dust lanes normally reveal locations of these contact points, and indeed some are seen close to the locations of the youngest YMCs at AZ$\sim80$ and 260 (Fig.~\ref{threecolor}) also corresponding to regions of highest nebular emission. The difference between the youngest clusters and the bar PA is similar to what \citet{ryder2010} find for HII regions in the nuclear ring of IC~4933.  Though it is difficult to exactly define the contact points in our case the mere presence of an age gradient is consistent with a dynamically (e.g.\ bar) induced cluster formation in the ring, a "pearls-on-a-string" scenario \citep{boker2008,seo2013} rather than localised gravitational instabilities resulting in a more random pattern (a "pop-corn model").  Also, any systematics in the age distribution indicate that multiple rotations, the rotational period of material at the ring radius is $\sim$30 Myr, have not washed out the signal confirming the generally young ages of clusters in NGC~2328.  Interestingly, according to models of \citet{seo2013}, merely finding an age gradient suggests the ring would have to be in a state of lower than average star formation rate (SFR).  Investigating how this fits with the time scales of cluster formation, and in particular {\em cluster formation efficiency} (CFE) which can be estimated using the newly derived MF and the SFR of the ring from our \ha\ image, will be done in our follow-up work on the SF characteristics of the whole galaxy.  We merely note here that CFE in the ring seems to be very high, much higher than in disks of ordinary SF galaxies.

\section{Summary}

Massive star clusters appear to be forming very efficiently in a low-luminosity early type galaxy NGC\,2328.  The clusters in a nuclear spiral/ring are young, possibly all $< 30$ Myr. 
The young massive clusters show an azimuthal age gradient, at least over large sections of the ring, suggesting bar driven gas flow at a not-too-high rate, a pearls-on-a-string cluster formation scenario. We observe a robust turn-over in the mass function of the clusters.
Assuming clusters are born with a power-law distribution, this turn-over indicates extremely rapid disruption.  If such cases are found to be typical, it would indicate the YMCs created in tight SF rings {\em do not survive to become old globular clusters}. 
 What in the environment is destroying them?  While the high density and shear are likely to be involved we are investigating the physical properties of the galaxy itself in more detail (V\"ais\"anen et al. in prep) to better understand the formation and dissolution mechanisms.

\acknowledgments

We are grateful to the referee, and to Stuart Ryder and Alexei Kniazev for useful comments and suggestions. 
PV and SB acknowledge support from the National Research Foundation of South Africa, and ZR from the SKA.
Some of the observations reported in this paper were obtained with the Southern African Large Telescope (SALT). Based on observations made with the NASA/ESA Hubble Space Telescope, and obtained from the Hubble Legacy Archive, which is a collaboration between the Space Telescope Science Institute (STScI/NASA), the Space Telescope European Coordinating Facility (ST-ECF/ESA) and the Canadian Astronomy Data Centre (CADC/NRC/CSA).

\end{document}